\documentclass[prl,aps,twocolumn,preprintnumbers,amssymb,nobibnotes,nofootinbib,superscriptaddress,raggedbottom,10pt]{revtex4-1}
\usepackage{hyperref,graphicx,array,multirow,color,amsmath,mathrsfs,titlesec,shuffle,tikz}
\usetikzlibrary{calc,decorations.markings}

\titleformat{\section}{\centering\normalsize\normalfont\bf}{\thesection}{0em}{}
\hypersetup{pdftitle={},pdfcreator={},linkcolor=[rgb]{0.15,0.35,0.75},colorlinks=true,citecolor=[rgb]{0.675,0,0.2},urlcolor=[rgb]{0.15,0.35,0.65}}

\newcommand{\fwbox}[2]{\text{\makebox[#1][c]{$\hspace{-150pt}\displaystyle#2\hspace{-150pt}$}}}
\newcommand{\fwboxL}[2]{\text{\makebox[#1][l]{$#2$}}}
\newcommand{\fwboxR}[2]{\text{\makebox[#1][r]{$#2$}}}

\newcommand{\eq}[1]{\vspace{-3.5pt}\begin{equation}\hspace{2pt}#1\hspace{-0pt}\vspace{-3.5pt}\end{equation}}
\newcommand{\eqL}[1]{\eq{\fwboxL{0pt}{\hspace{-115pt}#1}}}
\newcommand{\fig}[3]{\raisebox{#1}{\includegraphics[scale=#2]{#3}}}
\newcommand{\mi}{\raisebox{0.75pt}{\scalebox{0.75}{$\hspace{-1pt}\,-\,\hspace{-0.75pt}$}}}
\renewcommand{\pl}{\raisebox{0.75pt}{\scalebox{0.75}{$\hspace{-1pt}\,+\,\hspace{-0.75pt}$}}}

\newcommand{\equivR}{\fwbox{13.5pt}{\hspace{-0pt}\fwboxR{0pt}{\raisebox{0.47pt}{\hspace{1.45pt}:\hspace{-3pt}}}=\fwboxL{0pt}{}}}

\newcommand{\edgeA}{\text{{\normalsize${\color{hblue}a}$}}}\newcommand{\edgeB}{\text{{\normalsize${\color{hblue}b}$}}}\newcommand{\edgeC}{\text{{\normalsize${\color{hblue}c}$}}}\newcommand{\edgeD}{\text{{\normalsize${\color{hblue}d}$}}}\newcommand{\edgeE}{\text{{\normalsize${\color{hblue}e}$}}}\newcommand{\edgeF}{\text{{\normalsize${\color{hblue}f}$}}}\newcommand{\edgeG}{\text{{\normalsize${\color{hblue}g}$}}}\newcommand{\edgeH}{\text{{\normalsize${\color{hblue}h}$}}}
\newcommand{\br}[1]{\mathbin{\hspace{-1.5pt}\big[\hspace{-2.75pt}\big[#1\big]\hspace{-2.75pt}\big]\hspace{-1.5pt}}}
\definecolor{hblue}{rgb}{0,0,0.575}
\definecolor{hred}{rgb}{0.575,0.0,0.225}
\definecolor{hteal}{rgb}{0.0,0.545,0.7451}

\begin{document}
\title{\texorpdfstring{All-Multiplicity Non-Planar MHV Amplitudes in sYM at Two Loops\\[-12pt]~}{All-Multiplicity Non-Planar MHV Amplitudes in sYM at Two Loops}}
\author{Jacob~L.~Bourjaily}
\affiliation{Niels Bohr International Academy and Discovery Center, Niels Bohr Institute,\\University of Copenhagen, Blegdamsvej 17, DK-2100, Copenhagen \O, Denmark}
\affiliation{Center for the Fundamental Laws of Nature, Department of Physics,\\ Jefferson Physical Laboratory, Harvard University, Cambridge, MA 02138, USA}
\affiliation{Institute for Gravitation and the Cosmos, Department of Physics,\\Pennsylvania State University, University Park, PA 16892, USA}
\author{Enrico~Herrmann}
\affiliation{SLAC National Accelerator Laboratory, Stanford University, Stanford, CA 94039, USA}
\author{Cameron~Langer}
\affiliation{Center for Quantum Mathematics and Physics (QMAP),\\Department of Physics, University of California, Davis, CA 95616, USA\\~}
\author{Andrew~J.~McLeod}
\affiliation{Niels Bohr International Academy and Discovery Center, Niels Bohr Institute,\\University of Copenhagen, Blegdamsvej 17, DK-2100, Copenhagen \O, Denmark}
\author{Jaroslav~Trnka}
\affiliation{Center for Quantum Mathematics and Physics (QMAP),\\Department of Physics, University of California, Davis, CA 95616, USA\\~}

\begin{abstract}
We give a closed-form, prescriptive representation of all-multiplicity two-loop MHV amplitude integrands in fully-color-dressed (non-planar) maximally supersymmetric Yang-Mills theory.
\end{abstract}
\maketitle

\vspace{-12pt}\section{Introduction}\label{introduction_section}\vspace{-16pt}
%
Recent years have been witness to tremendous advances in our understanding of---and our ability to compute---scattering amplitudes in perturbative quantum field theory (see e.g.~\cite{Mangano:1990by,Dixon:1996wi,Elvang:2013cua,Henn:2014yza,Dixon:2015der,Cheung:2017pzi} and references therein). Perhaps the most impressive testament to these advances is found in the planar limit of maximally supersymmetric ($\mathcal{N}\!=\!4$) Yang-Mills theory (sYM) \cite{Brink:1976bc,Gliozzi:1976qd}. In this theory, loop integrands can be recursed to all orders \cite{ArkaniHamed:2010kv}, with local formulae known at all particle multiplicities through three loops \cite{Bourjaily:2013mma,Bourjaily:2015jna,Bourjaily:2017wjl} and for four particles through ten loops \cite{Bourjaily:2011hi,Bourjaily:2015bpz,Bourjaily:2016evz}. Integrated expressions are also known for six particles through seven loops~\cite{Caron-Huot:2019vjl,Caron-Huot:2019bsq}, and symbols are known for seven particles through four loops \cite{Dixon:2016nkn,Drummond:2018caf}. These computational triumphs only scratch the surface of the theoretical advances that have accompanied them (see for example~\cite{Drummond:2006rz,Beisert:2010jr,ArkaniHamed:2009dg,ArkaniHamed:2009sx,ArkaniHamed:2012nw,Arkani-Hamed:2013jha}); having access to this increasingly substantial compendium of concrete `data' has unquestionably fueled more general progress. 

In non-planar theories, considerably less data is available.~This is true even for the simplest quantum field theories, such as color-dressed (or `non-planar') sYM and maximal ($\mathcal{N}\!=\!8$) supergravity (`SUGRA'). In both theories, amplitude integrands are known for four particles through five loops \cite{Bern:1997nh,Bern:1998ug,Bern:2007hh,Bern:2010tq,Bern:2012uc,Bern:2017ucb}, and for five or six particles only through a modest two loops \cite{Carrasco:2011mn,Bern:2015ple,Bourjaily:2019iqr}. (Notably, the three-loop four-particle \cite{Henn:2016jdu} and two-loop five-particle amplitudes have also recently been integrated \cite{Abreu:2018aqd,Chicherin:2018old,Chicherin:2018yne,Chicherin:2019xeg,Abreu:2019rpt}.) It is known that the integrands for SUGRA amplitudes will involve terms with arbitrarily bad ultraviolet behavior (such as double poles at infinity) starting from seven particles~\cite{Bourjaily:2018omh}. In contrast, amplitude integrands in sYM are expected to be free of poles at infinity to all loop orders \cite{Arkani-Hamed:2014via,Bern:2014kca,Bern:2015ple}. Therefore, these amplitude integrands should be expressible in terms of an integrand basis with `triangle power counting' (a notion whose precise definition beyond the planar limit must wait until \cite{integrandBases}).

In this Letter, we show that this is indeed the case by presenting the first fully explicit, color-dressed, prescriptive representation of all-multiplicity MHV amplitude integrands in sYM at two loops. In the spirit of Parke and Taylor's original `guess' at tree-level \cite{Parke:1986gb} and similar guesses---later proven---at one \cite{Bern:1993qk,Bern:1994zx}, two \cite{Vergu:2009tu,ArkaniHamed:2010kv}, and three loops \cite{ArkaniHamed:2010gh,Bourjaily:2017wjl} in the planar sector, we have checked that our result smoothly reproduces known results for four through six particles, and passes many non-trivial tests at higher multiplicity.

\vspace{-12pt}\section{All-Multiplicity MHV Amplitude Integrands}\label{results_section}\vspace{-14pt}

The two-loop MHV all-multiplicity sYM integrand representation we construct in this work is explicitly \emph{prescriptive} \cite{Bourjaily:2017wjl}: expressed in terms of a basis of integrands diagonalized with respect to a spanning set of field theory cuts (see e.g.~\cite{Feng:2012bm,OPP,Mastrolia:2012wftl,Mastrolia:2013kca,Abreu:2017xsl} for related work). Our basis consists of all two-loop integrands that have at most single poles at infinity. In terms of these, all integrands with more than $4L$ propagators at $L$ loops are reducible, making the system explicitly triangular in cuts (and hence easy to diagonalize). It is worth pointing out that our setup leads to a surprisingly small number of relevant integrand basis elements in comparison to (for arbitrary $n$) the infinite number of Feynman diagrams required in traditional field theory methods (or even for BCJ \cite{BCJ}, for example). Because individual integrands have support on poles at infinite loop momentum, the cancellation of these residues at infinity for sYM amplitudes \cite{Bourjaily:2018omh} amounts to a non-trivial consistency check.

\begin{table*}[t!]
\vspace{-8pt}$$\fwbox{86pt}{\begin{array}{@{}c@{}}\text{\textbf{kissing boxes}}\\[-4pt]\fwbox{90pt}{\fig{-32.5pt}{0.9}{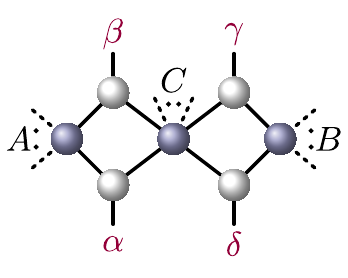}}\\[-3pt]
\fwboxR{5pt}{\text{kb}}\!\big[{\color{hred}\alpha},\!{\color{hred}\beta},\!{\color{hred}\gamma},\!{\color{hred}\delta},\!A,\!B,\!C\big]
\end{array}}
\fwbox{85.5pt}{\begin{array}{@{}c@{}}\text{\textbf{pentabox}}\\[-4pt]\fwbox{75pt}{\fig{-32.5pt}{0.9}{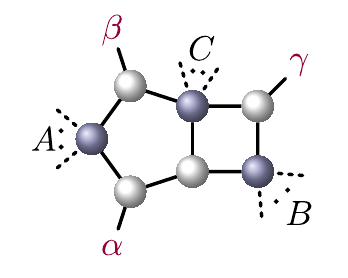}}\\[-3pt]
\fwboxR{5pt}{\text{pb}}\!\big[{\color{hred}\alpha},\!{\color{hred}\beta},\!{\color{hred}\gamma},\!A,\!B,\!C\big]
\end{array}}
\fwbox{85.5pt}{\begin{array}{@{}c@{}}\text{\textbf{hexabox A}}\\[-4pt]\fwbox{80pt}{\fig{-32.5pt}{0.9}{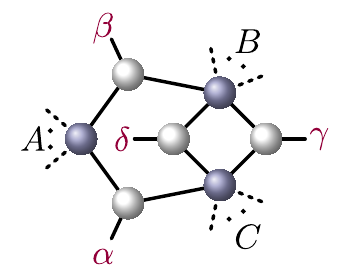}}\\[-3pt]
\fwboxR{10pt}{\text{hbA}}\!\big[{\color{hred}\alpha},\!{\color{hred}\beta},\!{\color{hred}\gamma},\!{\color{hred}\delta},\!A,\!B,\!C\big]
\end{array}}
\fwbox{86pt}{\begin{array}{@{}c@{}}\text{\textbf{hexabox B}}\\[-4pt]\fwbox{80pt}{\fig{-32.5pt}{0.9}{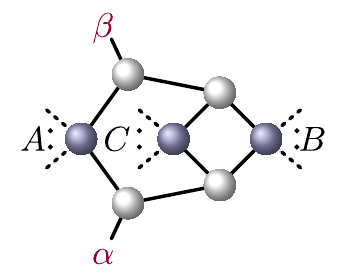}}\\[-3pt]
\fwboxR{10pt}{\text{hbB}}\!\big[{\color{hred}\alpha},{\color{hred}\beta},A,B,C\big]
\end{array}}
\fwbox{85.5pt}{\begin{array}{@{}c@{}}\text{\textbf{double pent.\ A}}\\[-4pt]\fwbox{70pt}{\fig{-32.5pt}{0.9}{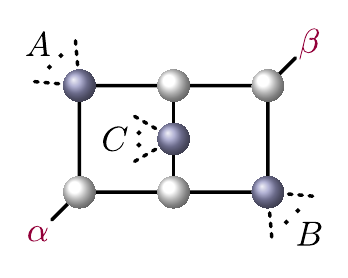}}\\[-3pt]
\fwboxR{10pt}{\text{dpA}}\!\big[{\color{hred}\alpha},\!{\color{hred}\beta},\!A,\!B,\!C\big]
\end{array}}
\fwbox{85.5pt}{\begin{array}{@{}c@{}}\text{\textbf{double pent.\ B}}\\[-4pt]\fwbox{70pt}{\fig{-32.5pt}{0.9}{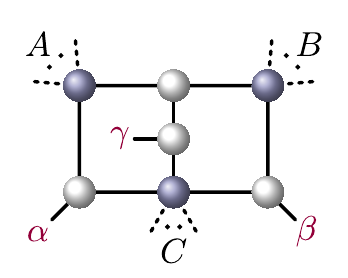}}\\[-3pt]
\fwboxR{10pt}{\text{dpB}}\!\big[{\color{hred}\alpha},\!{\color{hred}\beta},\!{\color{hred}\gamma},\!A,\!B,\!C\big]
\end{array}}\vspace{-14pt}$$
\caption{$\fwboxL{462.5pt}{\text{The six topology groups of two-loop leading singularities of MHV amplitudes. Explicit formulae are given in \mbox{\cite{Bourjaily:2019iqr}}.}}$\label{ls_table} \vspace{-10pt}}
\end{table*}

As our basis is diagonal in a spanning set of cuts, each integrand's coefficient is simply a residue of field theory---in our case, always a leading singularity (or zero). Thus, our representation takes the simple form:
\eqL{\mathcal{A}_{n}^{\text{MHV},(2\text{-loop})}=\;\;\hspace{-8pt}\sum_{\substack{\text{inequivalent}\\\text{leading singularities $\mathfrak{f}$}\\[-5pt]~}}\hspace{-10pt}\mathfrak{f}\,\times\,\mathcal{I}_{\mathfrak{f}}\label{two_loop_mhv_amplitudes}\vspace{-20pt}}
where $\mathfrak{f}$ belongs to one of the six classes of (color-dressed) field theory leading singularities with MHV-helicity support given in \mbox{Table \ref{ls_table}}. The sum is over all distributions of external legs. Two leading singularities are considered \emph{equivalent} if they are isomorphic as helicity-decorated graphs. Helicity degrees for MHV/$\overline{\text{MHV}}$ are indicated in \mbox{Table \ref{ls_table}} by blue and white vertices, respectively.

The representation (\ref{two_loop_mhv_amplitudes}) is a sum over \emph{all} distinct leg distributions---including cases where the sets of legs $A,B,C$ attached to MHV vertices are allowed to be empty. Such leading singularities have the interpretation of a residue taken in a soft (and sometimes collinear) region, which sets the momentum flowing through the `doubled' propagator to zero. The numerators of integrands corresponding to such cases always become proportional to the `doubled' propagator, leaving us with an ordinary collection of Feynman propagators. For example, 
\vspace{-8pt}\eqL{\begin{split}&\hspace{-1.6cm}\fwbox{120pt}{\fig{-37pt}{1}{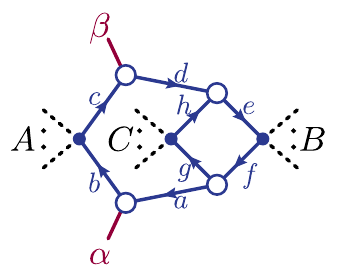}} \hspace{-14pt}\stackrel{C\to\emptyset}{\longrightarrow} \hspace{10pt}\fwbox{80pt}{\fig{-37pt}{1}{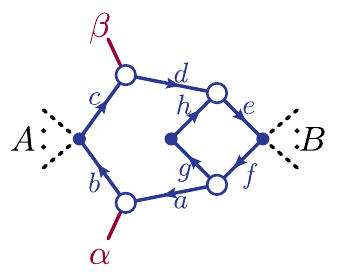}} \hspace{0pt}\\[-4pt]
\hspace{-8pt}\mathfrak{n}_{\text{hbB}}  = &\mi\!\br{{\color{hred}\alpha},\!\edgeB,\!\edgeC,\!{\color{hred}\beta}}\Big(\br{\edgeA,\!\edgeG\!,\!\edgeH,\!\edgeD} \mi\edgeA^2\edgeH^2\mi\edgeD^2\edgeG^2\!\Big)\\
\stackrel{C\to\emptyset}{\longrightarrow}&\mi\!\br{{\color{hred}\alpha},\!\edgeB,\!\edgeC,\!{\color{hred}\beta}}\Big(\!\mi(\edgeA\mi\edgeD)^2\edgeG^2\!\Big)\!=\!\mi\!\br{{\color{hred}\alpha},\!\edgeB,\!\edgeC,\!{\color{hred}\beta}}\Big(\!\mi p_B^2\edgeG^2\Big)\,,
\end{split}}
where $p_B^2\equivR\big(\sum_{b\in B}p_b\big)^2$. (Moreover, we can see that this numerator will vanish when the total momentum $B$ is massless (or empty).)

Notice that our instruction to sum over `all' distributions of legs for the figures in \mbox{Table \ref{ls_table}} seems to include cases with massless triangles or even bubbles (as for hexabox B when $B\!=\!C\!=\!\emptyset$); in all such cases, the corresponding integrand numerators either vanish, or the contributions cancel in sum.

The leading singularities in \mbox{Table \ref{ls_table}}, appearing as coefficients $\mathfrak{f}$ in (\ref{two_loop_mhv_amplitudes}), should be understood as fully-color-dressed on-shell functions in sYM. As such, every (tree-amplitude) vertex is fully (Bose-)symmetric. Without any reference to a particular gauge group (or trace decomposition), these factors can be defined concretely in terms of locally-(cyclically-)ordered MHV on-shell diagrams \cite{Arkani-Hamed:2014bca,Bourjaily:2016mnp} and graphs built out of (graphs of) structure constants of the type considered in \cite{DelDuca:1999rs} (see also \cite{Ochirov:2016ewn,Ochirov:2019mtf}). Explicit expressions for all these leading singularities were given in \mbox{Appendix B} of \cite{Bourjaily:2019iqr}. (These formulae are all smooth under taking any of the leg ranges $A,B,C$ to be empty, requiring no `cases' in their definitions.)

\newpage\vspace{-12pt}\subsection{Explicit Integrand Topologies and Numerators}\vspace{-14pt}
%
Attached to each leading singularity of \mbox{Table \ref{ls_table}}, we must construct an integrand that has unit support on the corresponding point in loop momentum space. Recall that when an MHV vertex in a leading singularity has no external legs attached to it, the corresponding residue is to be understood as the double constraint taking the momentum through that edge to be on-shell \emph{and soft}. Normalizing these integrands on their associated kinematic points is a good start, but is not sufficient to define our basis.

As stated above, our starting point is a complete basis of integrands (in four dimensions) with at most single poles at infinite loop momentum. In this space, there will be many integrands (numerator degrees of freedom) that can be normalized at points where no amplitudes in sYM have support. For example, integrands of the type shown in \mbox{Table \ref{contact_term}} are defined to have unit residue on a contour defined by cutting all seven propagators and symmetrically sending each loop to infinity.\footnote{When one or both of the leg ranges $A,B$ are empty, the numerator in \mbox{Table \ref{contact_term}} cancels the doubled propagator; the contour is then defined by starting from the `heptacut' that takes all propagators on-shell and for which the momentum flowing through any empty vertices is taken to be collinear to either ${\color{hred}\alpha}$ or ${\color{hred}\beta}$.} As all amplitude integrands in sYM should vanish at these points, the coefficients of these integrands in the sum (\ref{two_loop_mhv_amplitudes}) must be zero. Nevertheless, the entire (initially triangular) system of integrands must be diagonalized.

\begin{table}[t]\vspace{-24pt}$$\begin{array}{@{}c@{}}~\\[6pt]
\text{\textbf{`contact terms' (normalized at $\infty$)}}\\[-2pt]
\fwbox{120pt}{\fig{-37pt}{1}{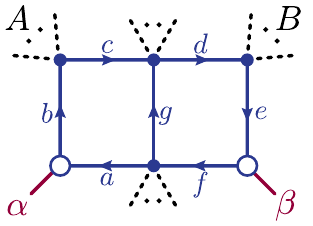}}\\[-3pt]
\fwboxR{0pt}{\phantom{\Big)}\mathfrak{n}_{\text{}}\equivR}\frac{1}{2}\br{{\color{hred}\alpha},\!\edgeB,\!\edgeC,\!\edgeD,\!\edgeE,\!{\color{hred}\beta}}
\end{array}\vspace{-18pt}$$
\caption{Basis integrands normalized at infinite loop momentum. In sYM, all these integrands have vanishing coefficients; nevertheless, all other integrands must be diagonalized with respect to these, accounting for the `contact terms' in the definition of integrands appearing in \mbox{Table \ref{integrand_table2}}. (Integrands in \mbox{Table \ref{integrand_table1}} are automatically diagonal with respect to these.) \\[-40pt]\label{contact_term}}\vspace{7pt}~
\end{table}

\begin{table*}[t]
\vspace{-4pt}$$
\fwbox{128pt}{\begin{array}{@{}c@{}}\text{\textbf{kissing boxes}}\\[-4pt]\fwbox{120pt}{\fig{-37pt}{1}{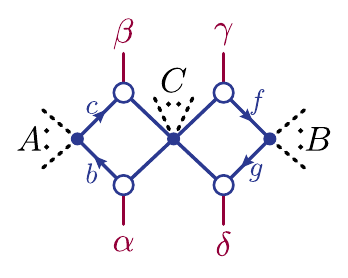}}\\[-3pt]
\fwboxR{0pt}{\mathcal{I}_\text{kb}\!}\big[{\color{hred}\alpha},{\color{hred}\beta},{\color{hred}\gamma},{\color{hred}\delta},A,B,C\big]\\[2pt]
\fwboxR{0pt}{\phantom{\Big)}\mathfrak{n}\equivR\!}\br{{\color{hred}\alpha},\!\edgeB,\!\edgeC,\!{\color{hred}\beta}}\br{{\color{hred}\gamma},\!\edgeF\!,\!\edgeG,\!{\color{hred}\delta}}\\\phantom{\Big)}\end{array}}
\fwbox{128pt}{\begin{array}{@{}c@{}}\text{\textbf{hexabox A}}\\[-4pt]\fwbox{120pt}{\fig{-37pt}{1}{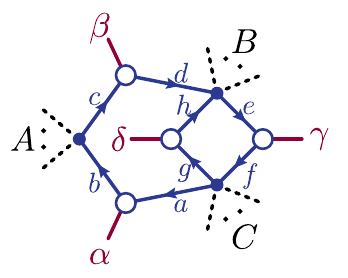}}\\[-3pt]
\fwboxR{0pt}{\mathcal{I}_{\text{hbA}}\!}\big[{\color{hred}\alpha},{\color{hred}\beta},{\color{hred}\gamma},{\color{hred}\delta},A,B,C\big]\\[2pt]
\fwboxR{0pt}{\phantom{\Big)}\mathfrak{n}\equivR\!}\br{{\color{hred}\alpha},\!\edgeB,\!\edgeC,\!{\color{hred}\beta}}\br{{\color{hred}\gamma},\!\edgeF\!,\!\edgeG,\!{\color{hred}\delta}}\\\phantom{\Big)}\end{array}}
\fwbox{128pt}{\hspace{10pt}\begin{array}{@{}c@{}}\text{\textbf{hexabox B}}\\[-4pt]\fwbox{120pt}{\fig{-37pt}{1}{hexaboxB_int}}\\[-3pt]
\fwboxR{0pt}{\mathcal{I}_{\text{hbB}}\!}\big[{\color{hred}\alpha},{\color{hred}\beta},A,B,C\big]\\[2pt]
\fwboxR{0pt}{\mathfrak{n}\equivR\!}\!-\!\br{{\color{hred}\alpha},\!\edgeB,\!\edgeC,\!{\color{hred}\beta}}\Big(\br{\edgeA,\!\edgeG\!,\!\edgeH,\!\edgeD}\\
\hspace{40pt}\mi\edgeA^2\edgeH^2\mi\edgeD^2\edgeG^2\Big)\end{array}}
\fwbox{128pt}{\begin{array}{@{}c@{}}\text{\textbf{double pentagon B}}\\[-4pt]\fwbox{120pt}{\fig{-37pt}{1}{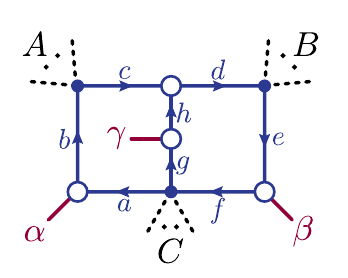}}\\[-3pt]
\fwboxR{0pt}{\mathcal{I}_{\text{dpB}}\!}\big[{\color{hred}\alpha},{\color{hred}\beta},{\color{hred}\gamma},A,B,C\big]\\[2pt]
\fwboxR{0pt}{\mathfrak{n}\equivR\!}\frac{1}{2}\Big(\!\br{{\color{hred}\alpha},\!\edgeB,\!\edgeC,\!\edgeD,\!\edgeE,\!{\color{hred}\beta},\!\edgeG,\!\edgeH}\\\hspace{8.05pt}-\br{{\color{hred}\alpha},\!\edgeB,\!\edgeC,\!\edgeD,\!\edgeE,\!{\color{hred}\beta},\!\edgeH,\!\edgeG}\!\Big)\end{array}}
\vspace{-18pt}$$
\caption{$\fwboxL{462.5pt}{\text{Integrand topologies with numerators that are smooth under all degenerations to empty leg-ranges.}}$\label{integrand_table1} \vspace{-10pt}}
\end{table*}

\begin{table}[t!]
\vspace{-10pt}$$
\fwbox{128pt}{\hspace{-10pt}\begin{array}{@{}c@{}}\text{\textbf{pentabox}}\\[-4pt]\fwbox{120pt}{\fig{-37pt}{1}{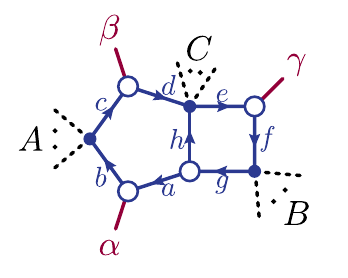}}\\[-3pt]
\fwboxR{0pt}{\mathcal{I}_\text{pb}\!}\big[{\color{hred}\alpha},{\color{hred}\beta},{\color{hred}\gamma},A,B,C\big]\\[2pt]
\fwboxR{0pt}{\phantom{\Big)}\mathfrak{n}_{\text{}}\equivR\!}\mi\!\br{{\color{hred}\alpha},\!\edgeB,\!\edgeC,\!{\color{hred}\beta}}\br{{\color{hred}\gamma},\!\edgeF\!,\!\edgeG,\!\edgeA}\\
\fwboxR{0pt}{\phantom{\Big)}}\hspace{0.pt}\pl\frac{1}{2}\br{{\color{hred}\alpha},\!\edgeB,\!\edgeC,\!{\color{hred}\beta},\!{\color{hred}\gamma},\!\edgeF\!,\!\edgeG,\!\edgeA}\fwboxL{0pt}{+\mathfrak{n}_{\text{pb}}^c\phantom{\Big)}}\\[0pt]
\end{array}}
\fwbox{120pt}{\hspace{-10pt}\begin{array}{@{}c@{}}\text{\textbf{double pentagon A}}\\[-4pt]\fwbox{120pt}{\fig{-37pt}{1}{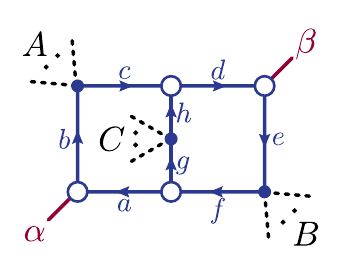}}\\[-3pt]
\fwboxR{0pt}{\mathcal{I}_\text{dpA}\!}\big[{\color{hred}\alpha},{\color{hred}\beta},A,B,C\big]\\[2pt]
\fwboxR{0pt}{\phantom{\Big)}\mathfrak{n}_{\text{}}\equivR\!}\mi\,\br{{\color{hred}\alpha},\!\edgeB,\!\edgeC,\!\edgeH,\!\edgeG,\!\edgeF,\!\edgeE,\!{\color{hred}\beta}}\\\hspace{-8pt}\fwboxL{57.5pt}{\pl\mathfrak{n}_{\text{dpA}}^c}\phantom{\Big)}\end{array}}\vspace{-6pt}$$\vspace{-0pt}$\fwboxL{250pt}{\text{where}}$
\vspace{-10pt}$$\fwboxL{230pt}{\mathfrak{n}_{\text{pb\phantom{A}}}^c\equivR\!\hspace{-0.5pt}\fwboxL{100pt}{\!\frac{1}{2}\!\!\left\{\!\begin{array}{@{}l@{$\;\hspace{10pt}$}l@{}}
\fwboxL{96pt}{0^{\phantom{2}}\phantom{\edgeF^2}}&\text{if }A\!\neq\!\emptyset, B\!\neq\!\emptyset\\[2pt]
\edgeB^2\edgeD^2\br{\!{\color{hred}\gamma},\!\hspace{-0.25pt}\edgeF\!,\!\edgeG,\!{\color{hred}\alpha}}&\text{if }A\!=\!\emptyset, B\!\neq\!\emptyset\\[2pt]
\!\edgeF^2\!\edgeD^2\!\br{{\color{hred}\alpha},\!\edgeB,\!\edgeC,\!{\color{hred}\gamma}}\mi\edgeF^2\!\edgeC^2\br{{\color{hred}\alpha},\!\edgeB,\!\edgeD,\!{\color{hred}\gamma}\!}&\text{if }B\!=\!\emptyset\\
\end{array}\right.}}$$
$$\fwboxL{230pt}{\mathfrak{n}_{\text{dpA}}^c\hspace{-0.5pt}\equivR\!\fwboxL{100pt}{\!\frac{1}{2}\!\!\left\{\!\begin{array}{@{}l@{$\qquad\hspace{9.55pt}$}l@{}}
\fwboxL{96pt}{0^{\phantom{2}}\phantom{\edgeF^2}}&\text{if }A\!\neq\!\emptyset, B\!\neq\!\emptyset\\[2pt]
\edgeB^2(\edgeG^2\pl\edgeH^2)\hspace{0.3pt}\br{\hspace{-0.3pt}{\color{hred}\beta},\!\edgeE,\!\hspace{-0.5pt}\edgeF\!,\!{\color{hred}\alpha}}&\text{if }A\!=\!\emptyset, B\!\neq\!\emptyset\\[2pt]\edgeE^2(\edgeG^2\pl\edgeH^2)\br{{\color{hred}\alpha},\!\edgeB,\!\edgeC,\!{\color{hred}\beta}}&\text{if }B\!=\!\emptyset\\
\end{array}\right.}}$$\vspace{-10pt}
\caption{Integrand topologies with numerators that have contact terms that change when some leg-ranges are empty.\label{integrand_table2}}\vspace{-5pt}
\end{table}

Besides points where general sYM amplitude integrands vanish, we take advantage of the particular simplicity of MHV amplitudes. Focusing on the simplest helicity configuration allows us to eliminate further integrand degrees of freedom by normalizing them on residues where MHV integrands have to vanish due to helicity selection rules. Note, however, that this procedure does not eliminate all integrand topologies without MHV support (see discussion below).  

It turns out that the `leading' (non-contact\footnote{To be clear, we define \emph{contact terms} as factors in the numerator proportional to one or more inverse propagators of the graph.}) terms of the integrand numerators for six-particle amplitudes in \cite{Bourjaily:2019iqr} are automatically diagonal with respect to themselves. However, the na\"ive numerators for two classes of integrands---the pentaboxes and double pentagons of type A---have support on the cuts defining the integrands of \mbox{Table \ref{contact_term}}. As such, diagonalization with respect to these `contact terms' results in some changes with respect to the na\"ive numerators. Taking into account these minor rotations in the basis, we obtain the form of our answer. 

(It is worth pointing out that while the numerator of hexabox B in \mbox{Table \ref{integrand_table1}} appears to have contact terms, these \emph{should not be viewed as contact terms}: they are fully fixed by graph symmetries, power-counting, integral purity, and chirality.)

The resulting integrand basis we find is summarized in \mbox{Tables \ref{integrand_table1}--\ref{integrand_table2}}. In \mbox{Table \ref{integrand_table1}} we list all the numerators, which are defined irrespective of whether or not any of the leg ranges $A,B,C$ are empty; in \mbox{Table \ref{integrand_table2}}, we give expressions for the numerators of the pentabox and double pentagon A integrands, which require contact terms depending on whether the leg ranges $A$ and/or $B$ are empty. 

These numerators are expressed in the notation
\eq{\br{a_1,a_2,\cdots,c_1,c_2}\equivR\!\Big[(a_1\!\cdot\!a_2)^{\alpha}_{\phantom{\alpha}\beta}\cdots(c_1\!\cdot\!c_2)^{\gamma}_{\phantom{\gamma}\alpha}\!\Big]\,,\label{definition_of_br}}
where $(a_1\!\cdot\!a_2)^{\alpha}_{\phantom{\alpha}\beta}\equivR a_1^{\alpha\,\dot{\alpha}}\epsilon_{\dot{\alpha}\dot{\gamma}}a_2^{\dot{\gamma}\gamma}\epsilon_{\gamma\beta}$ and $a^{\alpha\dot\alpha}\equivR a^{\mu}\sigma_{\mu}^{\alpha\dot{\alpha}}$ are `$2\!\times\!2$' four-momenta, defined via the Pauli matrices. (Our `$\br{\cdots}$' may be more familiar if written as `$\mathrm{tr}_+[\cdots]$'.)

\vspace{-12pt}\subsection{Cancellation of Calabi-Yau Cut Components}\vspace{-14pt}
%
As explained in \mbox{\cite{Bourjaily:2019iqr}}, local integrand representations of MHV amplitudes require terms that individually have support on elliptic and K3 (`Calabi-Yau') sub-topologies \cite{Bourjaily:2017bsb,Bourjaily:2018ycu,Bourjaily:2018yfy,Bourjaily:2019hmc}. This is despite the fact that these amplitudes are unquestionably polylogarithmic. The easiest way to see this is to notice that after cutting the six propagators of the `tardigrade' integral \cite{Bourjaily:2018yfy}
\vspace{-4pt}\eq{\fig{-39pt}{1}{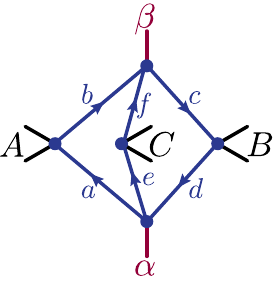}\label{k3_tardigrade}\vspace{-4pt}}
there is no helicity flow consistent with MHV. The same argument applies when any \emph{one} of $A,B,C$ becomes massless---the case first relevant to seven particles, where the integral becomes elliptic. Thus, we must ensure that our integrand representation vanishes identically on the two-dimensional surface defined by cutting the six propagators $\edgeA,\ldots,\edgeF$ of (\ref{k3_tardigrade}). 

There are nine integrals in (\ref{two_loop_mhv_amplitudes}) that have the K3 (\ref{k3_tardigrade}) as a sub-topology: six distributions of legs corresponding to double pentagon A, and three corresponding to hexabox B. It turns out that the Calabi-Yau six-cut cancels nontrivially via three sets of three-term identities. To illustrate this cancellation in more detail, consider the three integrands participating in one of the identities:
\vspace{-4pt}\eqL{\hspace{-10pt}\fig{-32.5pt}{0.9}{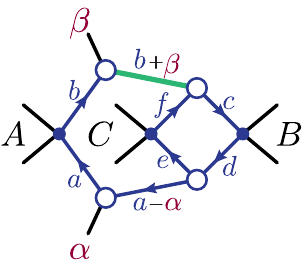}\hspace{-4pt}\fig{-32.5pt}{0.9}{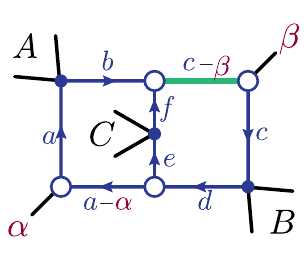}\hspace{-2pt}\fig{-32.5pt}{0.9}{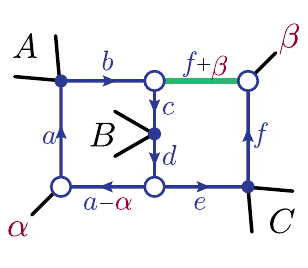}\label{jacob_related_ints}}
These integrands have the following leading-singularity coefficients:
\vspace{-4pt}\eqL{\hspace{-10pt}\fig{-32.5pt}{0.9}{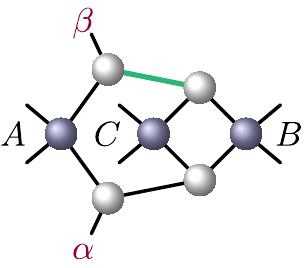}\hspace{-4pt}\fig{-32.5pt}{0.9}{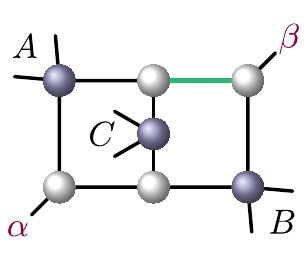}\hspace{-2pt}\fig{-32.5pt}{0.9}{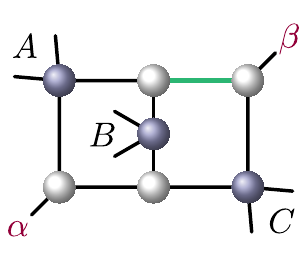}\vspace{-6pt}
\label{k3_identity_ls}}
which share seven propagators, six of which are isomorphic to those in (\ref{k3_tardigrade}), and differ only by the propagator highlighted in green. The fact that the leading singularities add to zero is a consequence of the Jacobi relation (the color-dressed merge-and-expand relation for on-shell functions \cite{ArkaniHamed:2012nw,Herrmann:2016qea}). This relation alone does not ensure the cancellation of the K3 cut;  however, it turns out that \emph{all nine} integrands evaluate to the \emph{same function} (up to a sign) of the two remaining degrees of freedom on the six-cut (\ref{k3_tardigrade}): 
\eqL{\begin{split}&
\hspace{-9pt}\left.\frac{\br{{\color{hred}\alpha},\!\edgeA,\!\edgeB,\!{\color{hred}\beta}}\Big(\!\br{\edgeA\mi{\color{hred}\alpha},\!\edgeE,\!\edgeF,\!\edgeB\pl{\color{hred}\beta}}\mi(\edgeA-{\color{hred}\alpha})^2\edgeF^2\mi(\edgeB\pl{\color{hred}\beta})^2\edgeE^2\!\Big)}{(\edgeA\mi{\color{hred}\alpha})^2(\edgeB\pl{\color{hred}\beta})^2}\!\right|_{\text{cut}}\\
&\;\fwboxR{0pt}{=}
\left.-\,\frac{\br{{\color{hred}\alpha},\!\edgeA,\!\edgeB,\!\edgeF,\!\edgeE,\!\edgeD,\!\edgeC,\!{\color{hred}\beta}}
}{(\edgeA-{\color{hred}\alpha})^2(\edgeC-{\color{hred}\beta})^2}\!\right|_{\text{cut}}\hspace{-5.95pt}=\left.-\,\frac{\br{{\color{hred}\alpha},\!\edgeA,\!\edgeB,\!\edgeC,\!\edgeD,\!\edgeE,\!\edgeF,\!{\color{hred}\beta}}
}{(\edgeA-{\color{hred}\alpha})^2(\edgeF+{\color{hred}\beta})^2}\!\right|_{\text{cut}}\hspace{-5.95pt}
\end{split}\label{equivalence_of_integrands_on_hexacut}}
where the first line is minus the integrand for the hbA contribution. To make better sense of (\ref{equivalence_of_integrands_on_hexacut}), notice that we have labeled the integrands (\ref{jacob_related_ints}) decorating the leading singularities of (\ref{k3_identity_ls}) according to the cut specified in (\ref{k3_tardigrade}) and evaluated the numerators defined in \mbox{Tables \ref{integrand_table1}--\ref{integrand_table2}} in terms of these loop-momentum labels. 

\vspace{-12pt}\subsection{Further Consistency Checks}\vspace{-14pt}
%
Besides the cancellation of the Calabi-Yau six-cut, we have performed a number of nontrivial consistency checks of our new result. In particular, we have explicitly compared the two-loop four-, five-, and six-particle integrands to known results \cite{Bern:1997nh,Carrasco:2011mn,Bourjaily:2019iqr}. Furthermore, in the planar sector we compared our result to that obtained from loop-recursion relations \cite{ArkaniHamed:2010kv}. We did not check all eight-particle unitarity cuts, however we made sure that our answer passes a large number of five- and six-cut checks, which involve almost all integrand topologies of our answer. (No new integrand topologies appear beyond eight particles.) Matching these low cuts correctly constitutes a highly nontrivial consistency check on our result. 

\vspace{-12pt}\subsection{Infrared Divergences and Infrared Finiteness}\vspace{-12pt}
%
The structure of infrared divergences has played a major role in understanding fundamental properties of gauge theories, see e.g.~\cite{Weinberg:1965nx,Catani:1998bh,Bern:1998sc,Aybat:2006mz,Dixon:2008gr,Gardi:2009qi,Becher:2009cu,Feige:2014wja}. As with the representation found for six particles in \cite{Bourjaily:2019iqr}, the representation of MHV amplitudes in (\ref{two_loop_mhv_amplitudes}) manifests as much of the infrared structure of the theory as possible. Specifically, all of the soft and collinear regions of loop-momentum space related to infrared divergences of amplitudes are matched manifestly, with coefficients that directly suggest something like exponentiation. By this, we refer to the fact that these leading singularities directly connect to lower-loop integrands (or trees) times products of factors that manifestly encode one-loop divergences. The precise sense in which this has something to say about how infrared divergences are organized in this representation---as compared with \cite{Catani:1998bh}, for example---remains to be explored.

One consequence of matching these infrared-divergent leading singularities directly (and diagonalizing our basis with respect to them) is that a large fraction of the terms in our basis are \emph{infrared finite}. In particular, only about half of the integrals required for six particles are infrared-divergent, and this fraction gets smaller at higher multiplicity. We strongly suspect that this feature will prove helpful in eventually finding analytic expressions for these amplitudes once regulated in some particular scheme.\\

It is worth emphasizing that the integrand we have constructed is strictly four-dimensional, and therefore not immediately suitable for dimensional regularization: in $4\!-\!2\epsilon$ dimensions, the integrand coefficients would change by terms of $\mathcal{O}(\epsilon)$, some of which survive as $\epsilon\to0$. (It would be interesting to know dimensionally-regularized integrands for these amplitudes, but constructing them would go beyond the scope of our present work.) That being said, the infrared structure of these four-dimensional integrands can be studied without subtlety using a Higgs/mass regulator \cite{Alday:2009zm,Henn:2010bk}, as any $\mathcal{O}(m^2)\log(m^2)$ terms would vanish in the $m^2\to0$ limit. 

\vspace{-12pt}\section{Conclusions and Discussion}\label{conclusions_section}\vspace{-14pt}
%
At tree-level and one loop, all amplitudes are built from terms that are planar with respect to some ordering. In this Letter we have given the first all-multiplicity formula for \emph{genuinely} non-planar scattering amplitude integrands. Our strategy avoided any reference to any particular gauge group, and required no choice of loop-momentum labels or routing. As such, we have demonstrated the power of prescriptive unitarity beyond the planar limit, opening the door to many future applications, including a better understanding of the structure of perturbative quantum field theory. \\

In the ancillary files for this work, we have prepared {\sc Mathematica} code making use of (\ref{two_loop_mhv_amplitudes}). Specifically, we provide documented functions to decompose, evaluate and expand into color-traces all leading singularities in \mbox{Table \ref{ls_table}}; represent and evaluate each of the integrands in \mbox{Tables \ref{integrand_table1}--\ref{integrand_table2}}; and to generate permutation-class representatives of each term appearing in (\ref{two_loop_mhv_amplitudes}) (together with the entire permutation sums) for arbitrary multiplicity. Useful tools for working with amplitudes in {\sc Mathematica} can be found in \cite{Bourjaily:2010wh,Bourjaily:2012gy}.

\nopagebreak

\vspace{-12pt}\section{Acknowledgments}\vspace{-15pt}
This work was supported in part by the Danish National Research Foundation (DNRF91), a grant from the Villum Fonden, and a Starting Grant \mbox{(No.\ 757978)} from the European Research Council (JLB,AJM), and a Carlsberg Postdoctoral Fellowship (CF18-0641) (AJM). The research of JT and CL\ is supported in part by U.S. Department of Energy grant DE-SC0009999 and by the funds of University of California. EH\ is supported by the U.S. Department of Energy under contract DE-AC02-76SF00515. Finally, JLB, EH, and AJM are grateful for the hospitality of the Aspen Center for Physics, which is supported by National Science Foundation grant PHY-1607611, and also to the Harvard Center for Mathematical Sciences and Applications. 

\vspace{-10pt}
\providecommand{\href}[2]{#2}\begingroup\raggedright\endgroup

\end{document}